\documentclass[letterpaper, 10pt, conference]{ieeeconf}
\IEEEoverridecommandlockouts
\usepackage{cite}
\usepackage{amsmath,amssymb,amsfonts}
\usepackage{algorithmic}
\usepackage{tabularx}
\usepackage{graphicx}
\usepackage{textcomp}
\usepackage{xcolor}
\usepackage{url} 
\usepackage{multirow}
\let\labelindent\relax      
\usepackage{enumitem}       
\usepackage{booktabs}    
\usepackage{caption}     
\usepackage{threeparttable}
\begin{document}



\title{ADAS-TO: A Large-Scale Multimodal Naturalistic Dataset and Empirical Characterization of Human Takeovers during ADAS Engagement}

\author{%
    Yuhang Wang, Yiyao Xu, Jingran Sun, Hao Zhou\textsuperscript{*}%
    \thanks{%
    All authors are with the Department of Civil and Environmental Engineering, 
    University of South Florida, Tampa, FL 33620, USA
    (e-mail: \{yuhangw, yiyaoxu, jingransun, haozhou1\}@usf.edu).
    \newline\indent
    \textit{*\,Corresponding author: Hao Zhou. haozhou1@usf.edu}}%
}


\maketitle


\begin{abstract}
Takeovers remain a key safety vulnerability in production ADAS, yet existing public resources rarely provide takeover-centered, real-world data. We present ADAS-TO, the first large-scale naturalistic dataset dedicated to ADAS-to-manual transitions, containing 15,659 takeover-centered 20\,s clips from 327 drivers across 22 vehicle brands. Each clip synchronizes front-view video with CAN logs. Takeovers are defined as ADAS ON$\rightarrow$OFF transitions, with the primary trigger labeled as brake, steer, gas, mixed, or system disengagement. We further separate planned driver-initiated terminations (Ego) from forced takeovers (Non-ego) using a rule-based partition. While most events occur within conservative kinematic margins, we identify a long tail of 285 safety-critical cases. For these events, we combine kinematic screening with vision--language (VLM) annotation to attribute hazards and relate them to intervention dynamics. The resulting cross-modal analysis shows distinct kinematic signatures across traffic dynamics, infrastructure degradation, and adverse environments, and finds that in 59.3\% of critical cases, actionable visual cues emerge at least 3\,s before takeover, supporting the potential for semantics-aware early warning beyond late-stage kinematic triggers. The dataset is publicly released at \url{huggingface.co/datasets/HenryYHW/ADAS-TO}.
\end{abstract}

\section{Introduction}
\label{sec:introduction}


Advanced Driver-Assistance Systems (ADAS) are increasingly deployed in production vehicles \cite{yurtsever2020survey, nidamanuri2021progressive} to improve comfort and safety, yet the transition of control authority back to the driver remains a critical safety vulnerability \cite{MATSUMURO2020237}. Understanding the safety risk and driver behaviors to prevent a crash when the system reaches its operational limits is essential, but empirical research is currently constrained by data limitations. Studies relying on driving simulators \cite{8082802, ZEEB2015212} often lack the complexity and behavioral realism of real-world traffic \cite{KARIMI2025108068, jui2025need}. While naturalistic driving studies successfully capture this real-world environment, compiling large-scale, multi-brand data on ADAS transitions is notoriously difficult. Existing large-scale datasets either lack kinematic Controller Area Network (CAN) bus logs \cite{caesar2020nuscenes, ettinger2021large} or lack the semantic context to explain \textit{why} a failure occurred \cite{dingus2016driver}. Conversely, studies that have complete data type are frequently constrained by small sizes \cite{fridman2019advanced}, manual annotation bottlenecks, or limitation to specific OEM platforms \cite{yang2023takeover}. Consequently, there remains a critical need for large-scale, diverse datasets that synchronize fine-grained physical metrics with visual semantic context to explain the environmental hazards triggering the driver's response.

To bridge the observability gap between vehicle kinematics and scene-level semantics, we present the ADAS-TO dataset. By synchronizing front-view dashcam video with high-frequency vehicle state and control signals, the dataset enables joint analysis of external driving context, driver interventions, and vehicle dynamics. This multimodal dataset can support failure-mode characterization and safety analysis of real-world ADAS takeovers at scale.

The main contributions of this paper are threefold:
\begin{itemize}
    \item \textbf{Large-Scale Multimodal Takeover Dataset:} We construct and release \textit{ADAS-TO}, comprising 15,659 clips of synchronized front-view video and CAN bus logs from 327 drivers across 22 vehicle brands, providing a solid empirical foundation for human-ADAS interaction research.
    
    \item \textbf{Cross-Modal Evaluation of Driver Responses:} We use intent partitioning, TTC/THW screening, and VLM hazard labels to connect scene context with driver interventions. We observe hazard-specific kinematic responses, including preemptive takeovers under adverse conditions and stronger coupled lateral/longitudinal disturbances in critical cases.
    
    \item \textbf{Potential for Proactive Early Warning:} We provide empirical evidence that semantic scene understanding can surface hazard-relevant cues earlier than purely kinematics-based triggers in critical takeover cases, highlighting the potential of multimodal perception for proactive ADAS alerting.
\end{itemize}

The remainder of this paper is organized as follows. Section II reviews related work. Section III details the data curation and dataset overview. Section IV presents the takeover analysis. Finally, Section V explores the potential applications of the dataset and our future work.

\section{Related Work}
\label{sec:related_work}

Most empirical takeover studies on production ADAS still build on takeover segments extracted from a small number of large naturalistic dataset. The MIT-AVT study instruments a limited fleet and records video, CAN, and GPS; however, data access is typically governed via consortium-style agreements rather than an openly downloadable, takeover-centered benchmark \cite{fridman2019advanced}. Using MIT-AVT, \emph{Takeover Context Matters} analyzes Super Cruise and Autopilot and performs hierarchical clustering on \emph{hundreds} of post-takeover kinematic traces, with context factors derived from video and GPS; the scope remains tied to two automation stacks and a comparatively small set of takeover cases \cite{yang2023takeover}. In contrast, ADAS-TO is organized as an event-centric release more data with synchronized front-view video CAN logs across more cars and more drivers, enabling broader cross-platform benchmarking.

Safety analysis and early-warning research for partial automation increasingly stresses fleet heterogeneity, supervision failures, and the limits of purely kinematic surrogate measures for explaining why interventions occur \cite{de2026cross}. Recent work on crash occurrence further suggests that combining semantic/contextual features with kinematics improves predictive power relative to any single modality \cite{wang2024context}. ADAS-TO directly supports this direction by aligning pre-takeover scene cues with high-rate intervention dynamics, allowing stratified analyses across drivers, vehicle brands/models, traffic regimes, and geography, and providing practical evaluation of multimodal warning strategies.

\section{Dataset Curation and Overview}                                
\label{sec:dataset}

\subsection{Data Sources}
\label{sec:data_collection}

\begin{figure}[!h]
    \centering
    \includegraphics[width=0.48\textwidth]{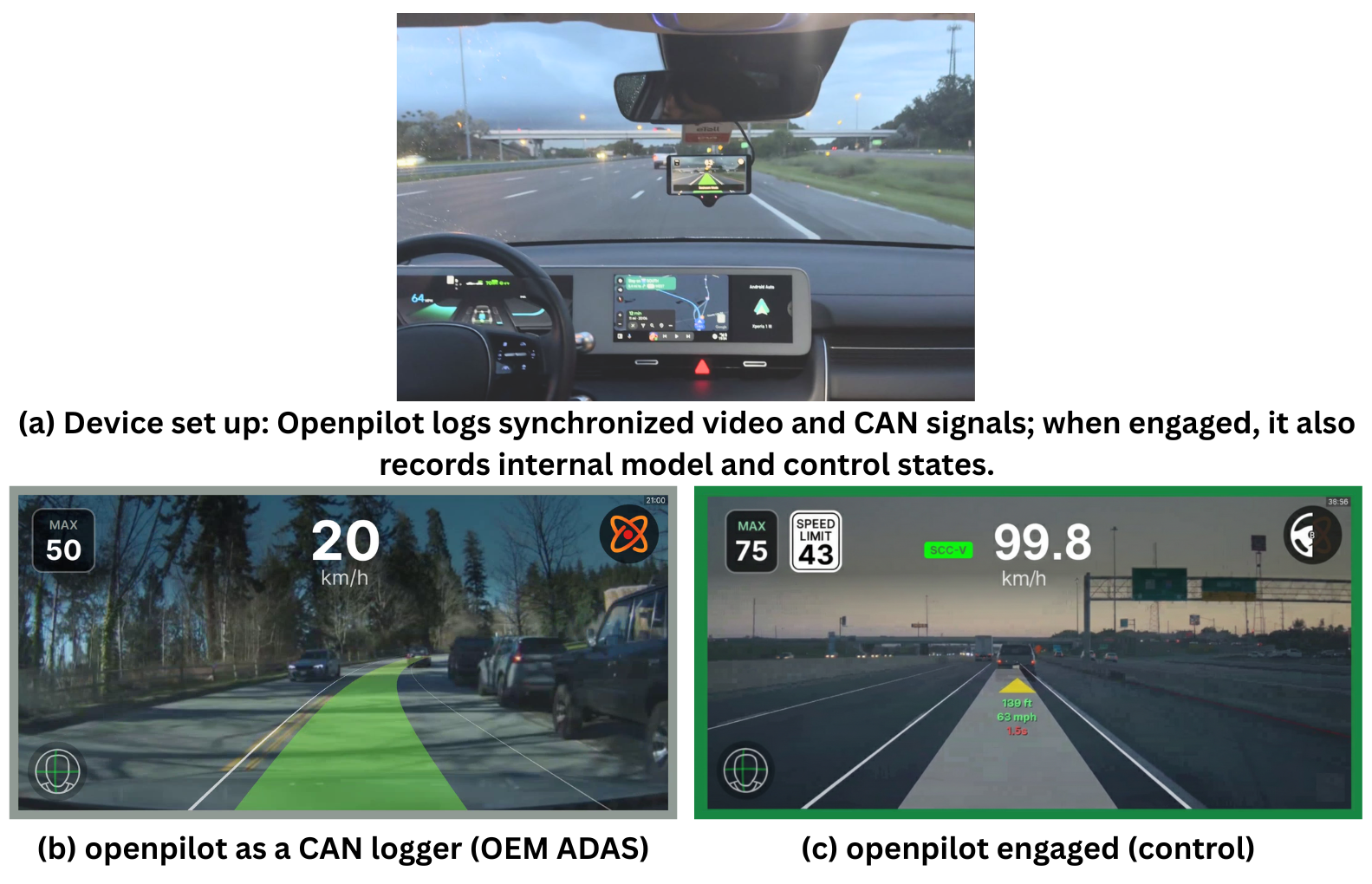}
    \caption{Role of OP in data collection.
    (a)~Device setup: Comma set up and OP performance 
    (b)~Passive logging under OEM ADAS: OP acts as an observer and captures vehicle dynamics and driver inputs around takeovers via CAN.
    (c) OP engaged (control): the system records its states and outputs from OP's driving model.}
    \label{fig:setup}
\end{figure}

\begin{figure*}[t]
    \centering
    \includegraphics[width=\textwidth]{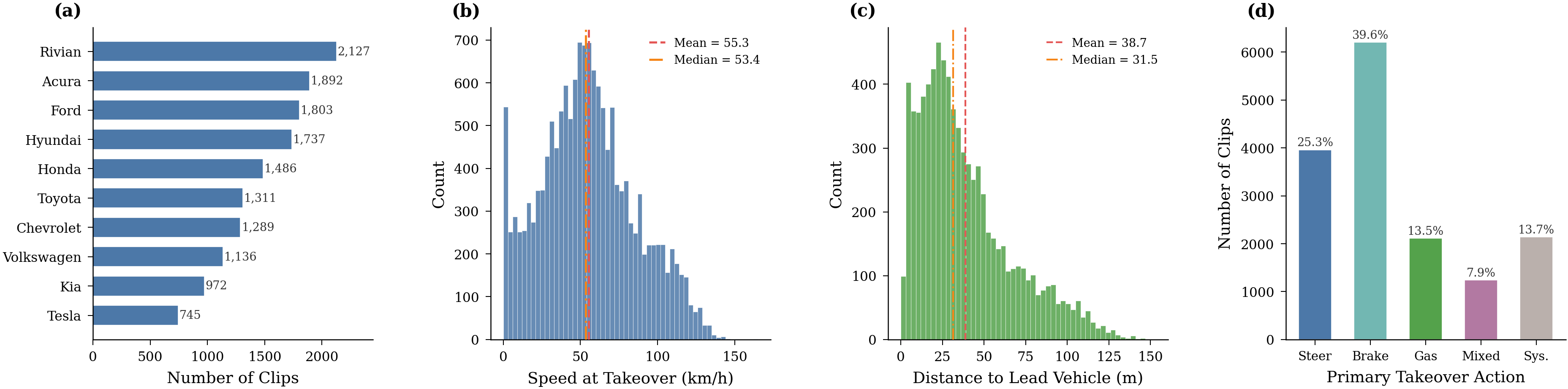}
    \caption{Dataset overview.
    (a)~Top-10 vehicle brands by clip count (out of 22 total brands).
    (b)~Distribution of vehicle speed at the takeover moment (dashed: mean; dash-dotted: median).
    (c)~Distribution of following distance to lead vehicle (when detected, 49.8\% of clips).
    (d)~Primary takeover action distribution; the first action applied to take over the vehicle.}
    \label{fig:dataset_overview}
\end{figure*}

The ADAS-TO dataset is constructed from naturalistic driving data recorded by market ADAS-equipped vehicles (Fig.~\ref{fig:dataset_overview}(a)) using comma 3/3X \cite{comma3x} devices mounted in the middle of the front window shield (Fig.~\ref{fig:setup}(a)) running \textit{openpilot} \cite{openpilot} (OP), an open-source driver assistance system that provides adaptive cruise control (ACC) and lane centering with driver monitoring. OP logs time-synchronized front-view video and CAN signals; when engaged for control, it additionally records internal model states and control commands.

\begin{figure}[t]
\centering
\includegraphics[width=\columnwidth]{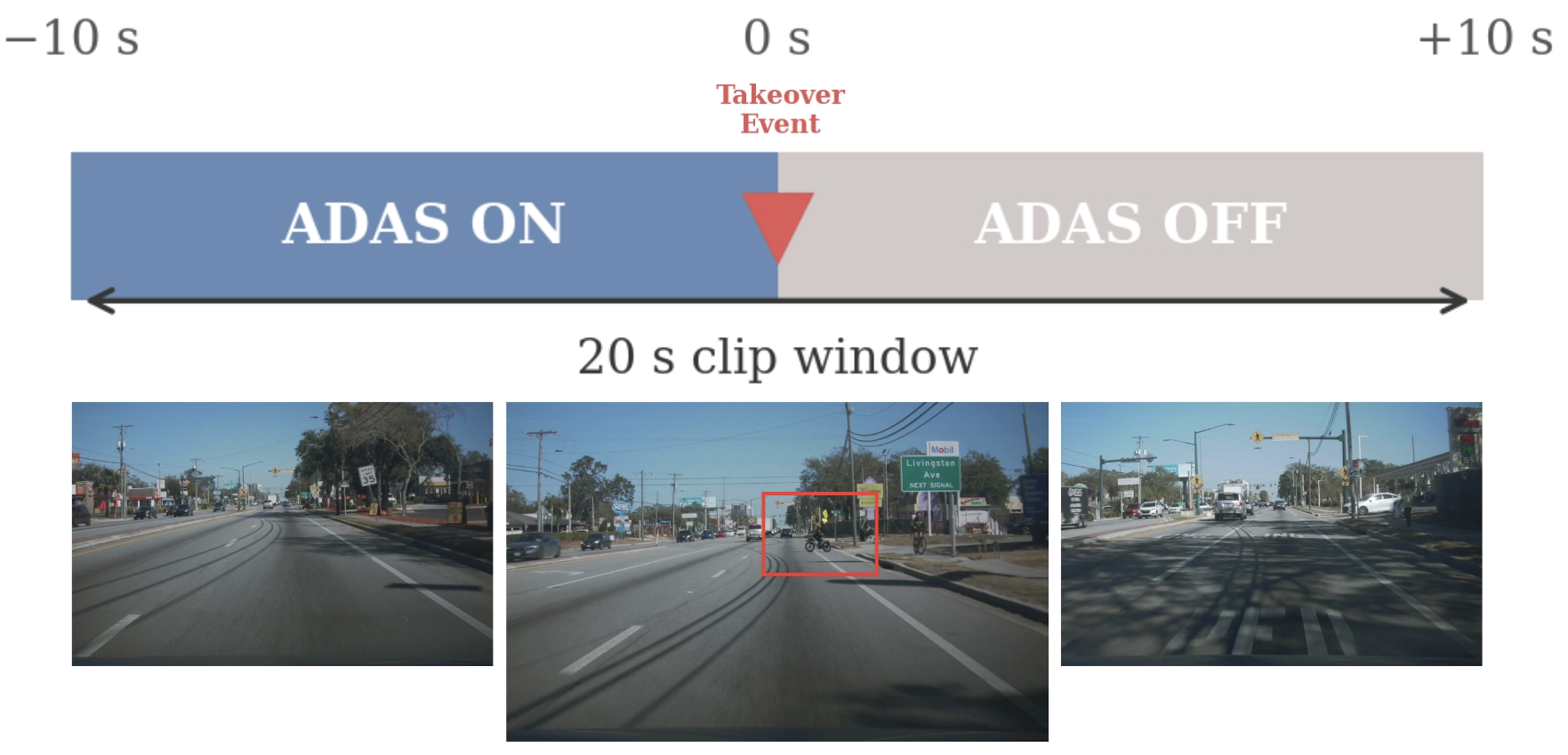}
\caption{Clip structure. Each 20\,s clip is centered on the takeover event ($t = 0$).}
\label{fig:clip_structure}
\end{figure}

Data were sourced from three complementary channels:
(1)~daily driving data from 4 experienced drivers
(2)~publicly shared driving data from the OpenLKA dataset~\cite{wang2025openlka}; and 
(3)~daily driving video and logs voluntarily contributed by autonomous driving community members operating similar autonomous driving kits, recruited through online forums and user
groups. 

\begin{figure}[t]
    \centering
    \includegraphics[width=\columnwidth]{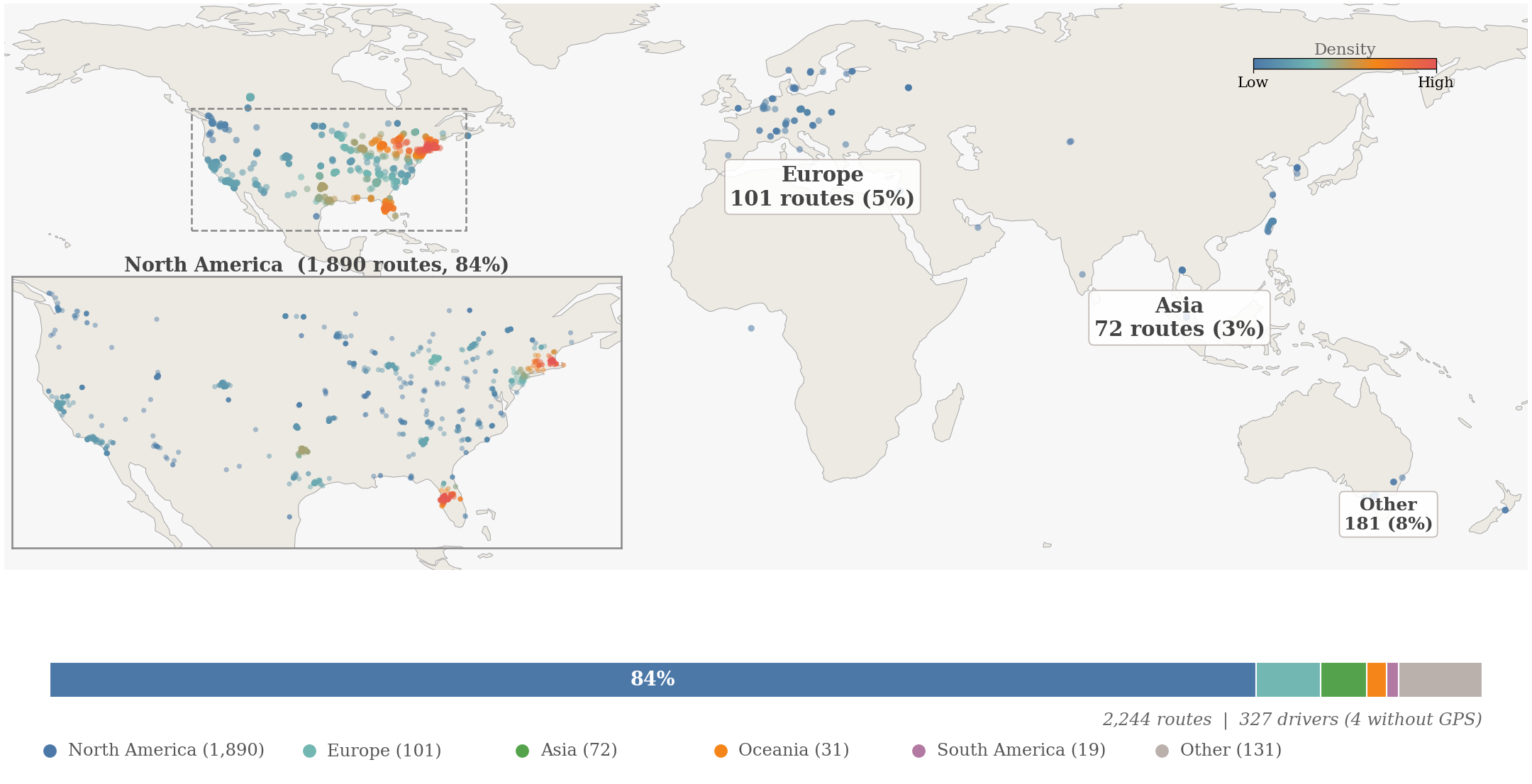}
    \caption{Geographic distribution of the 2{,}244 routes with GPS coordinates across 323 unique drivers with 4 don't want to share the location. North America accounts for     84.2\% of routes, followed by Europe     (4.5\%) and Asia (3.2\%).}
    \label{fig:geo_distribution}
\end{figure}

The dataset spans recordings from December~2019 to February~2026, covering OP releases from the v0.8.x series through v0.10.4. Geographically, the GPS-tagged routes are dominated by North America (84.2\%), with smaller shares from Europe (4.5\%) and Asia (3.2\%); the remaining routes (8.1\%) are distributed across other regions (Fig.~\ref{fig:geo_distribution}).

\subsection{Takeover Event Definition}
\label{sec:takeover_definition}

We define a \textit{takeover event} as a transition of the ADAS from the engaged (ON) state to the disengaged (OFF) state. The engagement indicator $E(t)$ is defined as the logical OR of two CAN-derived boolean flags:
OP ADAS control status and OEM ADAS control status, both signals are defined in OP’s log schema (cereal) \cite{cereal}. An ON$\to$OFF transition is registered when $E(t)$ has been continuously ON for at least $t_\text{on}=1.0$\,s and then continuously OFF for at least $t_\text{off}=1.0$\,s; brief ON/OFF glitches shorter than $0.5$\,s are merged to suppress spurious toggles.

The dataset includes multiple takeover mechanisms. Events may be system-initiated disengagements that require the driver to resume control, or driver-initiated takeovers. Driver-initiated takeovers may occur following system alerts, without explicit alerts based on the driver's judgment, or for discretionary reasons (e.g., comfort or preference).

For each detected event, we extract a 20-second clip centered on the transition time ($t=0$), spanning $[-10,+10]$\,s (Fig.~\ref{fig:clip_structure}). The symmetric $\pm 10$\,s window is chosen to capture the common pre-transition driving context and the immediate post-transition stabilization phase, while remaining compact enough for scalable model training and human inspection.

\subsection{Data Structure}
\label{sec:signals}

Each clip is stored as a self-contained directory with a front-view video (20\,fps), a metadata file, and eight time-synchronized CSV files covering the topics in Table~\ref{tab:signals}.
Signals are provided at either 10\,Hz (low-frequency \textit{qlog}, 61.7\% of clips) or 100\,Hz (high-frequency \textit{rlog}, 38.3\%).
For learning-based use, we recommend aligning all channels by \texttt{time} and resampling to a common rate; time-sensitive metrics should report sensitivity to the underlying log rate.

The dataset contains multiple OP releases (v0.8.x--v0.10.x; e.g., unified longitudinal MPC in v0.8.10 \cite{commaai_openpilot_v0810}, end-to-end longitudinal control in v0.9.0 \cite{commaai_openpilot_v090} Experimental Mode, configurable driving personality in v0.9.3 \cite{commaai_openpilot_v093}, and the World Model-based planner in v0.10.0 \cite{commaai_openpilot_v0100}).
Beyond OP-controlled driving, we also logs OEM ADAS engagement and driver/vehicle CAN signals in passive mode.
Across the full corpus (15{,}705 clips), 63.3\% involve OEM ADAS only, 16.0\% involve OP longitudinal control only, and 19.0\% have both active
(1.7\% other edge cases), with substantial variation across makes and models.
Per-route OP version and \texttt{git\_commit} are extracted from the log \texttt{InitData} message and recorded in the metadata, enabling source- and
version-stratified analyses.

\begin{table}[t]
\centering
\caption{Signal topics recorded per clip. Each clip contains eight time-synchronized CSV files.}
\label{tab:signals}
\footnotesize
\begin{tabular}{@{}ll@{}}
\toprule
\textbf{CSV File} & \textbf{Key Signals} \\
\midrule
\texttt{carState}         & Speed (\texttt{vEgo}), acceleration, steering angle/ \\
                 & torque, brake/gas/steering pressed, cruise state \\
\texttt{controlsState}    & ADAS enabled/active, curvature, desired curvature, \\
                 & set cruise speed, alert texts \\
\texttt{carControl}       & Lateral/longitudinal active, actuator commands \\
                 & (accel, torque, curvature) \\
\texttt{carOutput}        & Actuator outputs (accel, brake, gas, steer) \\
\texttt{drivingModelData} & Model-predicted curvature/acceleration, \\
                 & lane line detection probabilities \\
\texttt{radarState}       & Lead vehicle distance (\texttt{dRel}), relative speed, \\
                 & lead acceleration (primary \& secondary) \\
\texttt{longPlan}         & Target acceleration, lead flag, FCW, \\
                 & planned speed/acceleration trajectories \\
\texttt{accelerometer}    & Three-axis inertial acceleration \\
\bottomrule
\end{tabular}
\end{table}

To ensure data quality, we excluded 46 non-genuine clips (0.3\%) associated with system errors or invalid gear states, yielding a final dataset of \textbf{15,659} valid clips complete with all eight synchronized CSV files.

\begin{figure}[t]
\centering
\includegraphics[width=0.7\columnwidth]{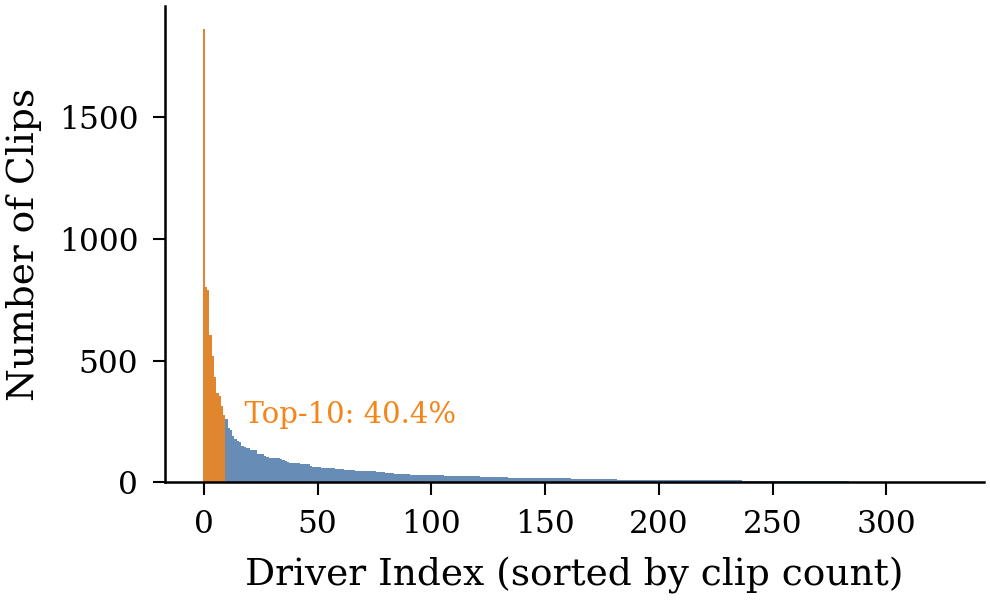}
\caption{Clips per driver. The distribution follows a long tail: the top-10 drivers (orange) contribute 40.4\% of all clips.}
\label{fig:clips_per_driver}
\end{figure}

\subsection{Takeover Action Classification}
\label{sec:actions}

To classify the modality of each takeover trigger, we examine three boolean CAN signals, steering torque, brake pedal and accelerator pedal, within a tight temporal window of $[-0.2, +0.5]$\,s relative to the disengagement instant ($t = 0$). This narrow window strictly isolates the immediate physical trigger while accounting for signal registration latency on the CAN bus.

Rather than applying a rigid hierarchy, we determine the \textit{primary action} based on temporal precedence: the first control modality to activate within this window is assigned as the primary trigger. If multiple control inputs register simultaneously at the first active timestamp, the event is classified as ``Mixed.'' If no driver override flags are detected within the window, the disengagement is labeled ``System'' (disengagement).

\subsection{Dataset Statistics}
\label{sec:statistics}

Table~\ref{tab:dataset_summary} and Fig.~\ref{fig:dataset_overview} summarize the key characteristics of the ADAS-TO dataset. The fleet comprises 163 car models from 22 brands.

\paragraph{Driver diversity and data skew}
The 15{,}659 clips originate from \textbf{327 unique drivers}.
The clip count per driver follows a long-tail distribution (Fig.~\ref{fig:clips_per_driver}): the top-10 drivers contribute 40.4\% of all clips, while the median driver contributes 15 clips and the mean is 48 (Fig.~\ref{fig:clips_per_driver}).
This within-driver correlation must be accounted for in any modeling task.
We recommend driver-disjoint splits as the default train/validation/test protocol to prevent within-driver information leakage.
For evaluating cross-platform generalization, a \textit{brand-disjoint} or \textit{car model-disjoint} out-of-distribution split is also supported by the metadata.

\paragraph{Speed conditions}
The vehicle speed at the moment of takeover has a mean of $54.9$\,km/h and a median of $53.3$\,km/h (Fig.~\ref{fig:dataset_overview}b).
We categorize clips into four speed regimes as a kinematic proxy:
standstill ($< 2$\,km/h, 3.6\%), low speed ($2$--$60$\,km/h, 56.1\%), medium speed ($60$--$100$\,km/h, 30.7\%), and high speed ($\geq 100$\,km/h, 9.7\%).

\paragraph{Takeover Actions Distribution}
As shown in Fig.~\ref{fig:dataset_overview}(d), Brake override is the most frequent primary action (39.6\%), followed by Steering (25.3\%), Gas (13.5\%), and Mixed inputs (7.9\%). System-initiated passive disengagements account for the remaining 13.7\%. This dominance of brake overrides indicates that drivers primarily rely on longitudinal interventions to mitigate immediate risks, establishing longitudinal dynamics as the focal point for safety-critical takeover analysis.

\paragraph{Lead vehicle}
Among clips where a lead vehicle is detected by radar (49.8\%), the mean following distance is $38.7$\,m (median $31.5$\,m), with 29.2\% of cases at close range ($< 20$\,m) (Fig.~\ref{fig:dataset_overview}c).
The lead vehicle flag is derived from radar; when no lead is reported, it may indicate either genuine absence of a preceding vehicle or a scenario where the radar does not detect one. The lead detection rate is consistent across both log types.

\begin{table}[t]
\centering
\caption{Summary statistics of the ADAS-TO dataset.}
\label{tab:dataset_summary}
\begingroup
\footnotesize
\setlength{\tabcolsep}{2.5pt}       
\renewcommand{\arraystretch}{0.95}  
\begin{tabular}{@{}p{0.56\columnwidth}@{\hspace{8pt}}r@{}}
\toprule
\textbf{Statistic} & \textbf{Value} \\
\midrule
Total takeover clips                 & 15{,}659 \\
Unique car models                    & 163 \\
Unique brands                        & 22 \\
Unique drivers (\texttt{dongle\_id}) & 327 \\
Total video duration                 & 87.0\,h \\
Clip duration                        & 20\,s ($\pm 10$\,s) \\
Collection period                    & Dec.\,2019--Feb.\,2026 \\
\midrule
\multicolumn{2}{@{}l}{\textit{Speed at takeover (km/h)}} \\
\quad Mean $\pm$ Std                 & $54.9 \pm 30.8$ \\
\quad Median [P5, P95]               & 53.3 [4.7, 111.1] \\
\midrule
\multicolumn{2}{@{}l}{\textit{Primary action (\%)}} \\
\quad Steering / Brake / Gas / Mixed / System& 25.3 / 39.6 / 13.5 / 7.9 / 13.7 \\
\midrule
\multicolumn{2}{@{}l}{\textit{Lead vehicle (49.8\%)}} \\
\quad Distance: Mean $\pm$ Std (m)   & $38.7 \pm 28.1$ \\
\midrule
\multicolumn{2}{@{}l}{\textit{Driver stats}} \\
\quad Mean / Median clips per driver & 47.9 / 15 \\
\quad Top-10 driver contribution     & 40.4\% \\
\midrule
\multicolumn{2}{@{}l}{\textit{Logging \& software}} \\
\quad 10\,Hz (qlog) / 100\,Hz (rlog) & 61.7\% / 38.3\% \\
\quad OP versions                    & v0.8.x--v0.10.x \\
\bottomrule
\end{tabular}
\endgroup
\end{table}

\section{Takeover Analysis}
\label{sec:take_over_analysis}

\subsection{Intent-Based Takeover Partition and Expert Validation}
\label{sec:intent_validation}

\begin{figure}[!h]                                           
    \centering
    \includegraphics[width=0.48\textwidth]{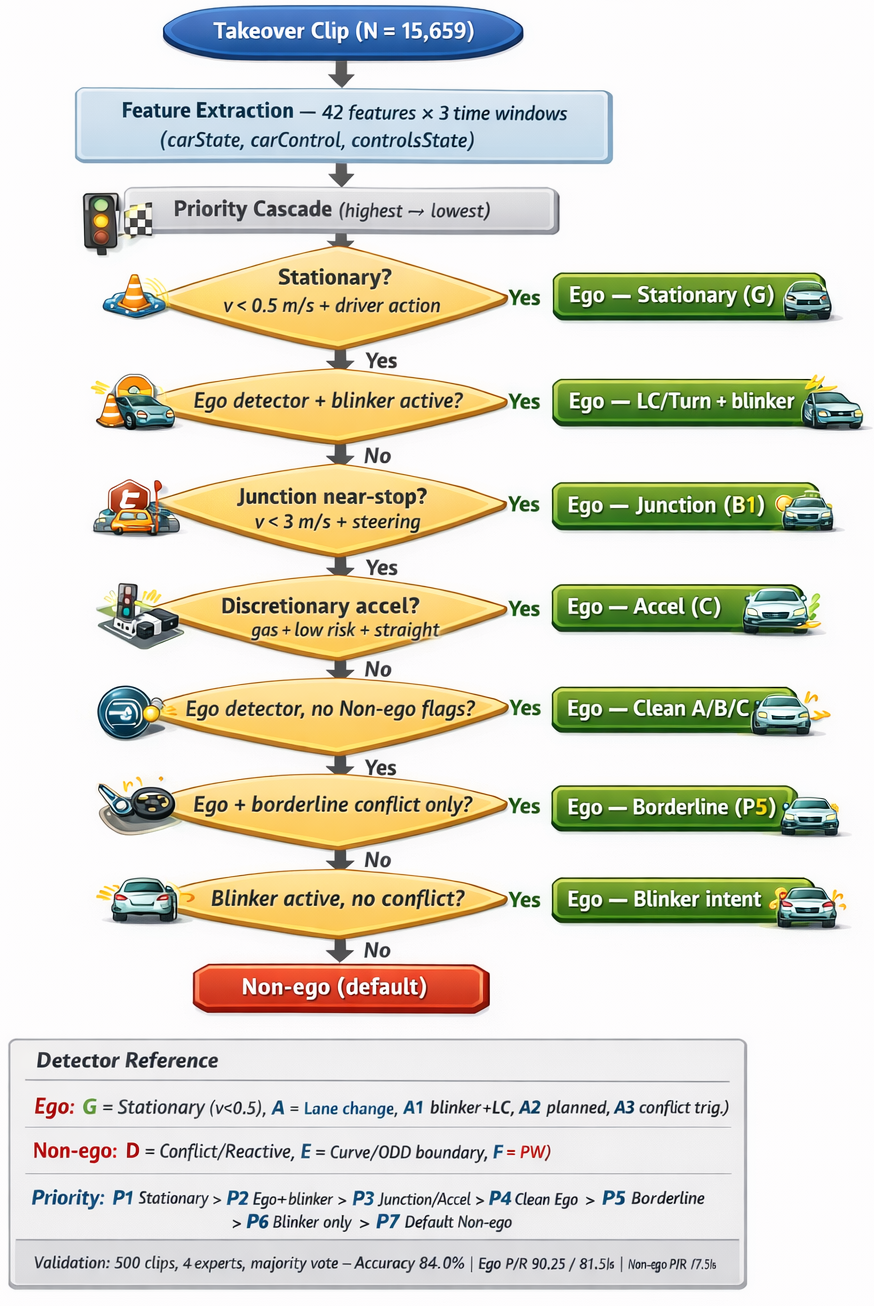}
\caption{Rule-based classification flowchart. Each takeover clip enters a seven-level priority cascade from top to bottom. At each decision node, the clip is tested against a specific detector condition; if matched, it is labeled \textit{Ego} (green) and exits the cascade. Clips not captured by any Ego rule are assigned the default \textit{Non-ego} label.}
    \label{fig:flowchart}
\end{figure}

Many takeovers in naturalistic logs are not safety-critical. To avoid mixing fundamentally different behaviors in downstream analysis, we partition takeovers into two intent classes:
\textbf{Ego} (driver-initiated, planned termination of ADAS, e.g., lane change/turn/roundabout/traffic light/stop sign/comfort stop)
and \textbf{Non-ego} (reactive or forced takeover due to system limits or external risk, e.g., proximity risk, sharp curvature, degraded
lane perception, or system-initiated disengagement).
We implement a rule-based classifier (Fig. \ref{fig:flowchart}) that combines drivers' post-takeover actions (e.g., blinker usage, steering/speed patterns)
with pre-takeover driving conditions (e.g., perceiving the sharp curve and low lane-confidence signals) to assign Ego and Non-ego labels for the full dataset. Subsequent analyses will focus more on safety analysis under Non-ego conditions rather than driver behavior under Ego conditions.

\paragraph{Expert audit protocol}
To evaluate the partition quality, we construct a stratified audit set of 500 clips (balanced by classifier output: 250 predicted Ego and
250 predicted Non-ego) spanning diverse vehicle platforms.
Four domain experts independently reviewed each clip using the 20\,s front-view video and synchronized CAN traces. Each annotator
assigned Ego vs. Non-ego by checking whether a deliberate maneuver appears after disengagement (e.g., turn, lane change, roundabout,
traffic control, or intentional stop). The final label is determined by majority vote; 2--2 splits are resolved by discussion to reach
consensus.

\begin{figure}[!h]
  \centering
  \includegraphics[width=\columnwidth]{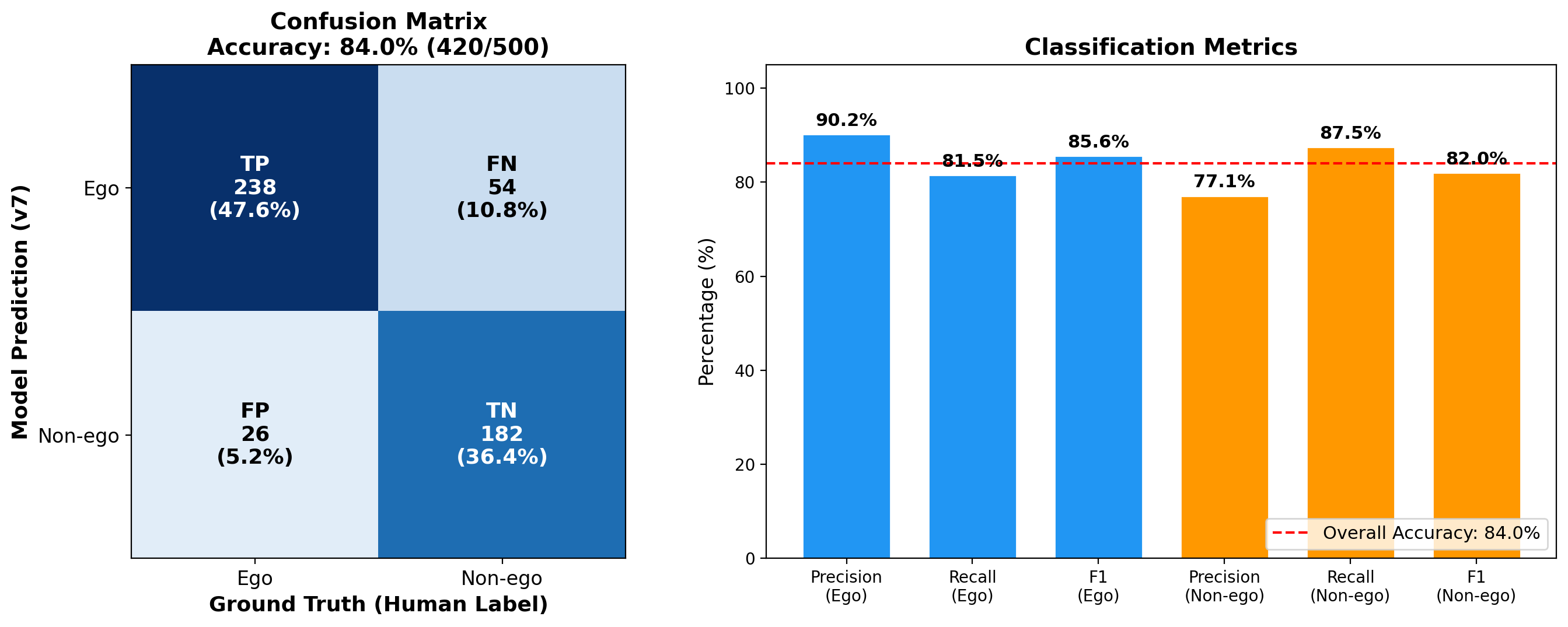}
  \caption{Expert validation of the Ego/Non-ego intent partition on a 500-clip audit set. Left: confusion matrix for the rule-based
  classifier. Right: per-class precision/recall/F1, with the dashed line indicating overall accuracy (84.0\%).}
  \label{fig:ego_nonego_validation}
\end{figure}

\paragraph{Audit results}
Fig.~\ref{fig:ego_nonego_validation} reports the confusion matrix and per-class metrics on the 500-clip audit set.
The rule-based partition achieves \textbf{84.0\%} accuracy (420/500), with \textbf{90.2\%} precision and \textbf{81.5\%} recall for Ego, and
\textbf{77.1\%} precision and \textbf{87.5\%} recall for Non-ego.
Most disagreements occur in borderline cases where an apparent maneuver co-occurs with contextual risk (e.g., lane change that appears
reactive to surrounding traffic), which is inherently ambiguous without direct observations of driver intent.

\begin{figure}[!h]
    \centering
    \includegraphics[width=\columnwidth]{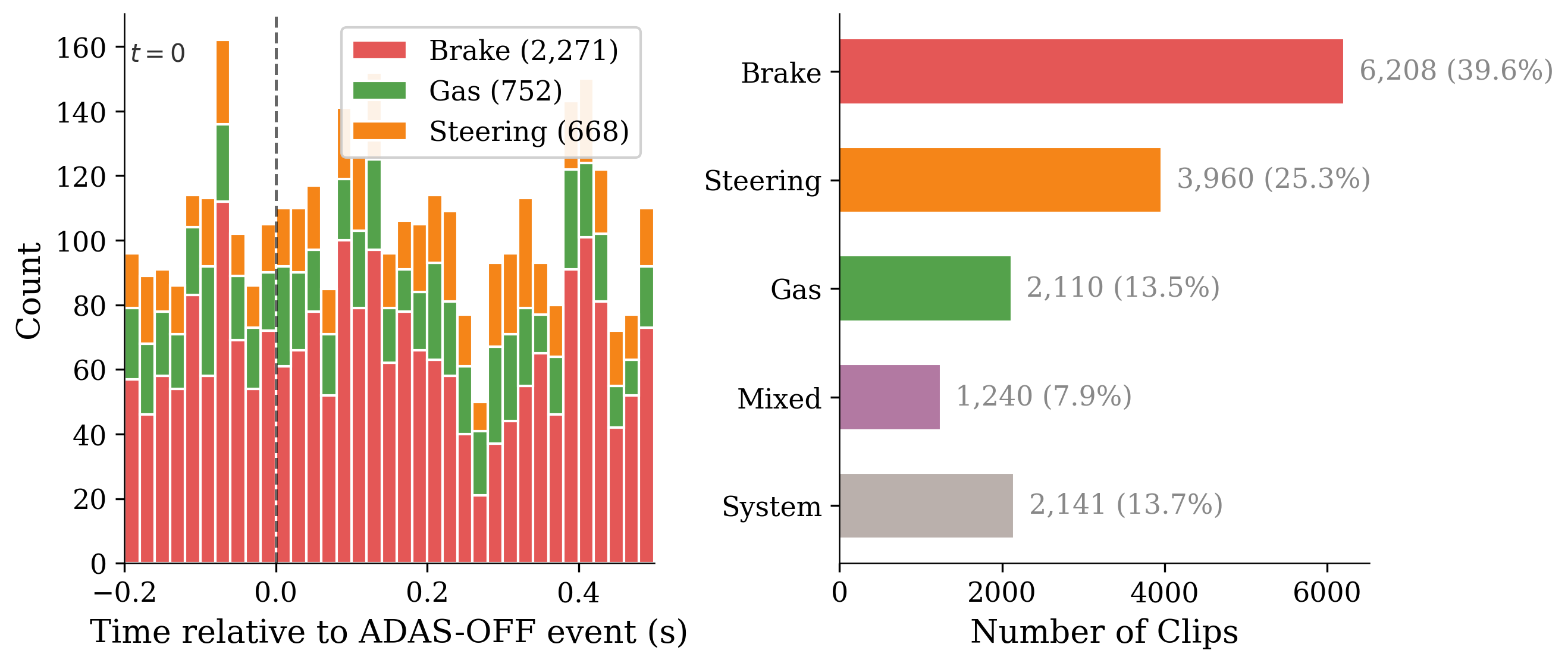}
    \caption{Takeover action sequence and modality distribution. (LEFT) The action onset distribution within the $[-0.2, +0.5]$\,s window illustrates the precise timing of control inputs relative to the ADAS disengagement ($t=0$). (RIGHT) The primary action modality highlights a clear longitudinal preference, with braking and accelerating take up nearly half of the data.}
    \label{fig:action_sequence}
\end{figure}

\subsection{Kinematic Profiling and Critical Long-Tail Extraction}
\label{sec:kinematic_profiling}

To characterize the physical execution of takeover events, we analyze the kinematic profiles within the $[-0.2, +0.5]$\,s window around the transition. As shown in Fig.~\ref{fig:action_sequence}, the temporal distribution of first-action onsets indicates a clear preference for longitudinal intervention: 39.6\% of takeovers are initiated via braking, compared to steering (25.3\%) and acceleration (13.5\%).

\begin{figure}[!h]
    \centering
    \includegraphics[width=0.95\columnwidth]{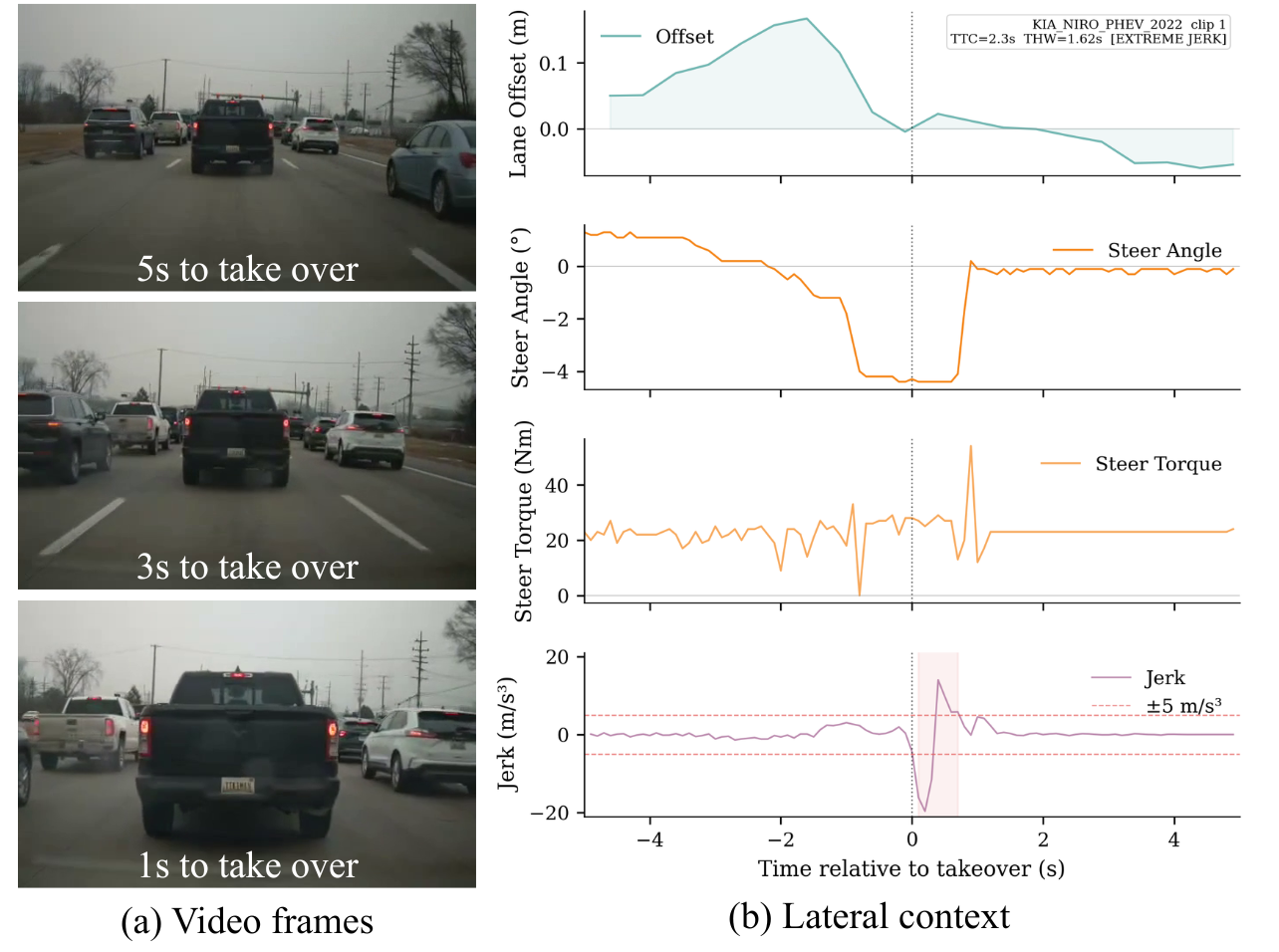}
    \caption{Example of a lateral takeover response. The subplots depict (a) the vehicle's deviation to lane center, alongside the driver interventions via (b) steering angle and (c) steering torque. }
    \label{fig:lateral_trajectory}
\end{figure}

\begin{figure}[!h]
    \centering
    \includegraphics[width=\columnwidth]{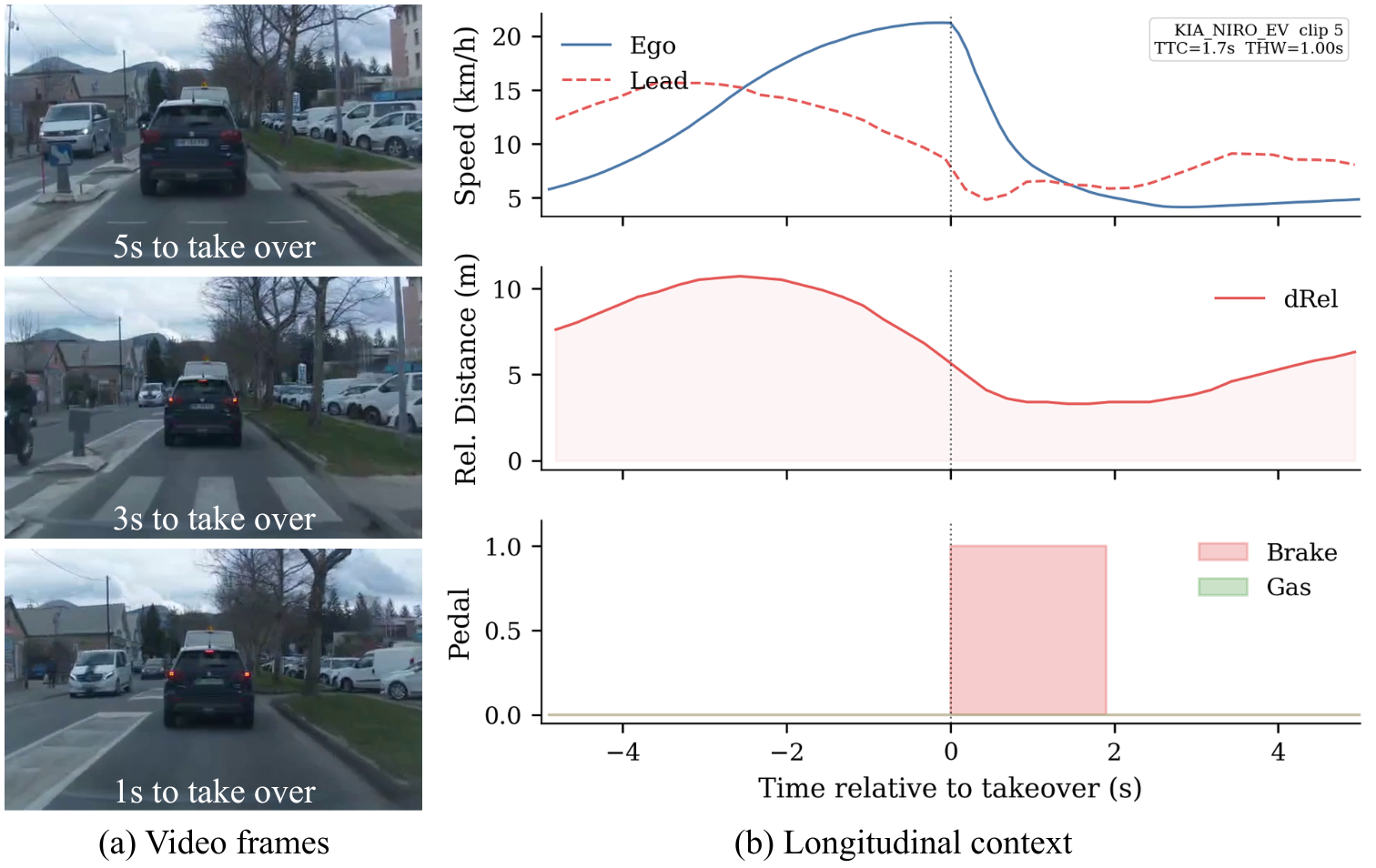}
    \caption{Longitudinal takeover context for a representative event. The subplots detail (a) the speed profiles of the ego and lead vehicles, (b) the relative distance to the lead vehicle, and (c) the driver's pedal inputs within the $[-5, +5]$\,s window around the takeover instant ($t=0$).}
    \label{fig:longitudinal_context}
\end{figure}

Despite this prevalence of brake-initiated overrides, an evaluation of the baseline metrics reveals that the vast majority of takeovers across both domains are preemptive rather than emergency responses. Laterally, the median absolute lane offset at the moment of takeover is merely 0.158\,m, representing a marginal deviation within a standard 3.27\,m lane (Fig.~\ref{fig:lateral_trajectory}). Longitudinally, for the 5,543 clips where a valid lead vehicle is consistently tracked ($d_{rel} > 2.0$\,m, closing speed $v_{rel} <  -  0.5$\,m/s), we compute the safety metrics under the standard constant-velocity assumption: $\text{TTC} = d_{rel} / |v_{rel}|$ and $\text{THW} = d_{rel} / v_{ego}$, where $v_{ego}$ is the ego vehicle speed. Based on these definitions, the median TTC \cite{kiefer2003forward} is 14.90\,s and the median THW \cite{vogel2003comparison} is 2.32\,s (Fig.~\ref{fig:longitudinal_context}). Both distributions confirm that the average intervention occurs well within safe operational boundaries.

\begin{figure}[!h]
    \centering
    \includegraphics[width=0.9\columnwidth]{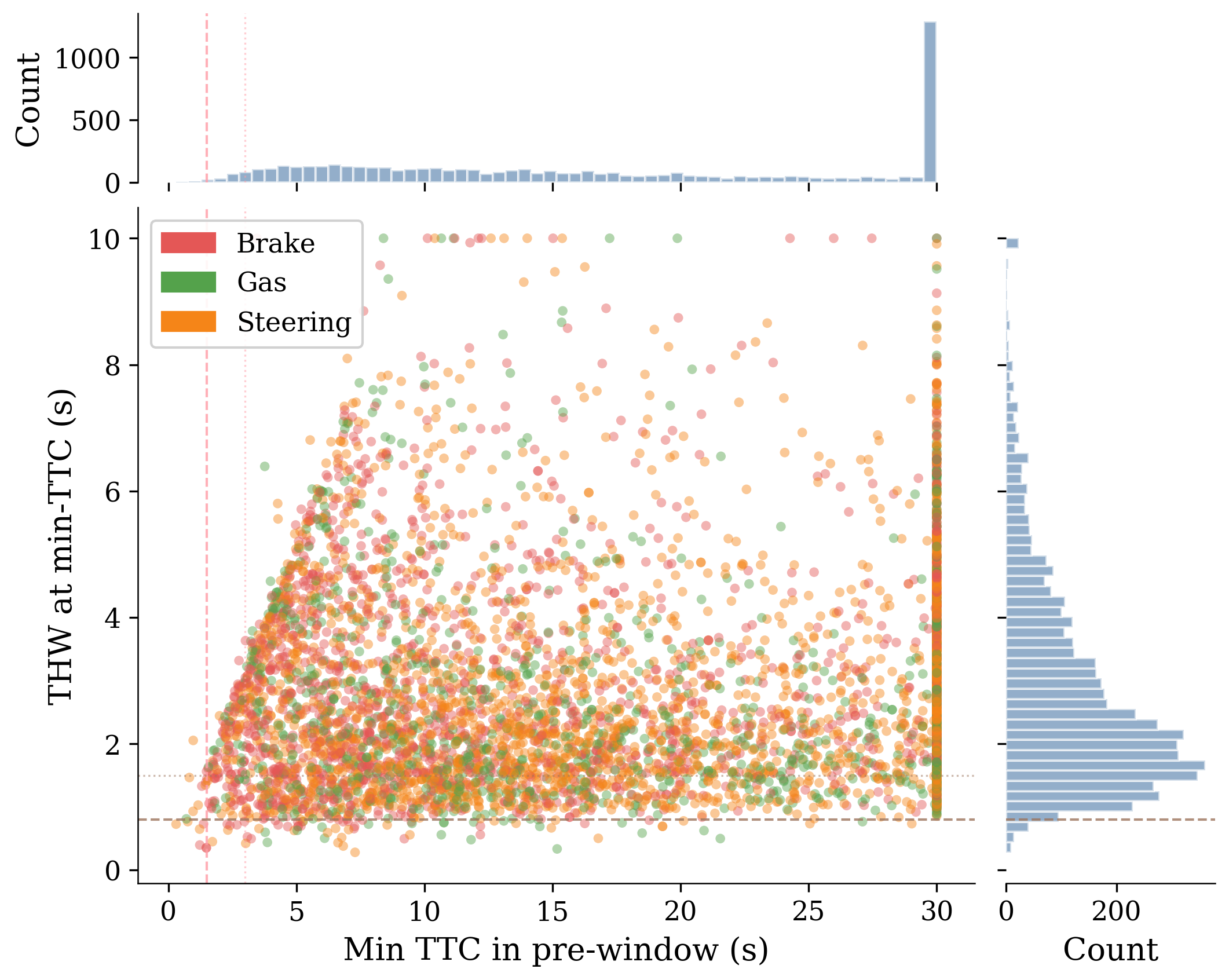}
    \caption{Distribution of TTC and THW at the moment of takeover initiation. A critical long-tail of high-risk events is distinctly visible.}
    \label{fig:ttc_thw_scatter}
\end{figure}

However, the critical danger lies in the extremities of these distributions, the ``long-tail.'' To systematically isolate genuine near-crash events from the massive subset of comfort, driven takeovers, we must establish an objective filter. While extreme lateral deviations are safety-critical, their absolute risk is highly context-dependent and ambiguous without environment perception. In contrast, longitudinal metrics provide an absolute, deterministic physical boundary for crash imminence. Therefore, we utilize established early-warning thresholds in the longitudinal domain ($\text{TTC} < 3.0$\,s or $\text{THW} < 0.8$\,s) as the primary sieve to extract high-risk events (Fig.~\ref{fig:ttc_thw_scatter}).

This thresholding successfully isolates 285 critical clips (173 with low TTC, 127 with low THW, and 15 intersecting). Crucially, post-takeover analysis within this critical subset reveals severe instability across \textit{both} control dimensions: 172 clips exhibit extreme longitudinal jerk ($\ge 5.0\,\text{m/s}^3$), accompanied by aggressive lateral evasive maneuvers in specific scenarios. This confirms that in true edge cases, drivers often struggle with both longitudinal and lateral vehicle stabilization.

While these kinematic metrics effectively quantify the intensity of the driver's reaction, they lack the semantic information necessary to explain \textit{why} the ADAS failed or what specific environmental hazards (e.g., cut-ins, faded lane markings) prompted the extreme response. To identify these underlying causal factors, we must integrate visual scene analysis into the evaluation pipeline.

\begin{figure*}[t] 
    \centering
    \includegraphics[width=\textwidth]{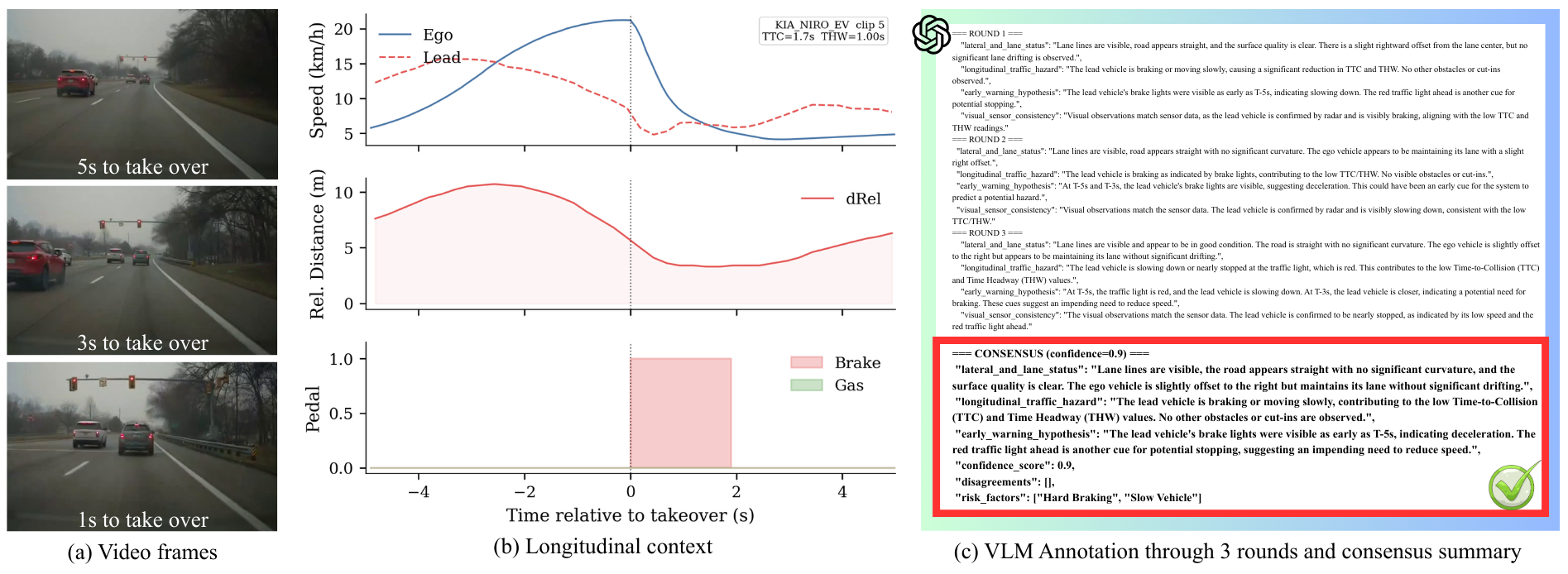} 
    \caption{Exemplar of the VLM-based semantic annotation pipeline. The model integrates three pre-takeover frames ($T-5$\,s, $T-3$\,s, $T-1$\,s) with synchronized sensor data. In this scenario, the VLM achieves a consensus confidence score of 0.9, correctly identifying early visual cues (brake lights and lead vehicle's low speed) that precede the low-TTC emergency.}
    \label{fig:vlm_example}
\end{figure*} 

\subsection{Vision-Language Semantic Annotation and Early Warning Analysis}
\label{sec:vlm_annotation}

To systematically attribute these 285 long-tail events to specific physical hazards, we employ a state-of-the-art VLM for automated spatiotemporal scene annotation.

To ensure the VLM accurately interprets the driving context, we formulate a multimodal prompt integrating both visual progression and ground-truth sensor data. For each critical clip, we extract three consecutive dashcam frames at $T-5$\,s, $T-3$\,s, and $T-1$\,s prior to the first driver intervention. These frames are provided to the VLM alongside synchronized sensor measurements (TTC, THW, ego speed, and lead vehicle distance) to establish a cross-modal grounding constraint. The model is tasked with analyzing the lateral lane status, longitudinal traffic hazards, and potential early visual cues that precede the critical event.

To mitigate VLM hallucinations and ensure annotation reliability, we implement a multi-round self-consistency framework. For each event, the VLM independently processes the prompt three times with a non-zero temperature ($\tau = 0.7$) to generate diverse interpretations. A subsequent consensus round ($\tau = 0.0$) acts as a meta-reviewer, synthesizing the three independent outputs to resolve minor discrepancies, outputting a final confidence score, and assigning standardized risk factor tags (e.g., ``Hard Braking'', ``Slow Vehicle''). 

Fig.~\ref{fig:vlm_example} illustrates this pipeline using a representative critical clip. The VLM consensus yields a high confidence score of 0.9 and assigns the risk factors ``Hard Braking'' and ``Slow Vehicle.'' The model describes the lane as straight with clear markings, noting a slight but stable rightward offset of the ego vehicle. The primary hazard is isolated to the longitudinal domain: a lead vehicle decelerating on approach to a red traffic light. Crucially, the VLM identifies that the red light and the lead vehicle's brake lights were visible as early as $T-5$\,s.

\subsection{Cross-Modal Analysis of Critical Takeovers}
\label{sec:cross_modal_analysis}

\begin{figure}[!h]
    \centering
    \includegraphics[width=\columnwidth]{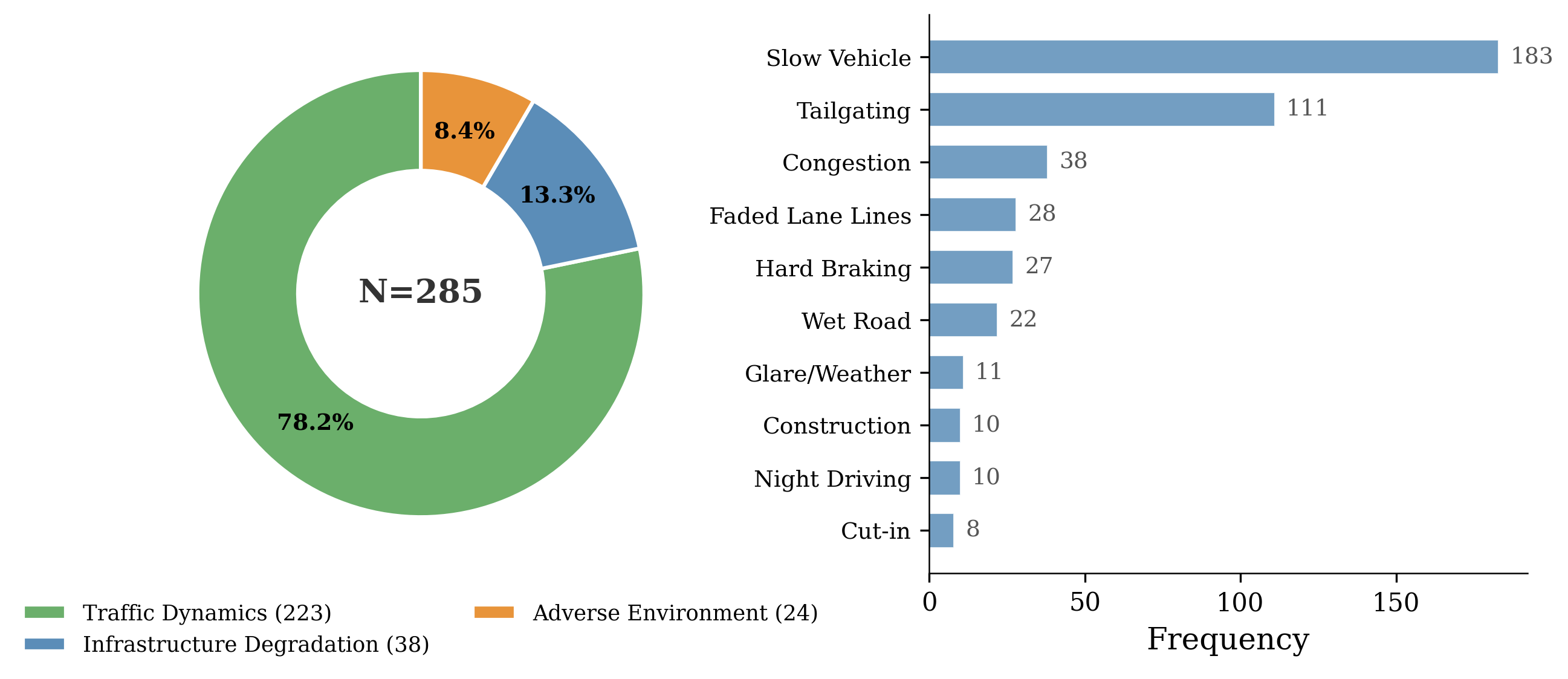}
    \caption{Semantic clustering of VLM-identified risk factors. The 285 critical takeover events are categorized into three macro-archetypes.}
    \label{fig:semantic_clustering}
\end{figure}

To evaluate how specific environmental hazards influence driver intervention behavior, we map the VLM-identified risk factors to the corresponding post-takeover kinematic data. We group the specific risk factors from the 285 critical clips into three macro-archetypes: Traffic Dynamics (e.g., slow vehicles, tailgating), Infrastructure Degradation (e.g., faded lane lines, construction zone), and Adverse Environment (e.g., wet roads, glare, night driving).

Fig.~\ref{fig:semantic_clustering} shows the distribution of these archetypes. Traffic Dynamics accounts for the majority of critical events (78.2\%, 223 clips), primarily driven by slow-moving vehicles and close following distances. Infrastructure Degradation constitutes 13.3\% (38 clips), and Adverse Environment conditions make up the remaining 8.4\% (24 clips).

By stratifying the post-takeover kinematic metrics by these semantic archetypes, we observe distinct intervention patterns for each failure mode (Fig.~\ref{fig:kinematic_signatures}):

\begin{figure}[!h]
    \centering
    \includegraphics[width=\columnwidth]{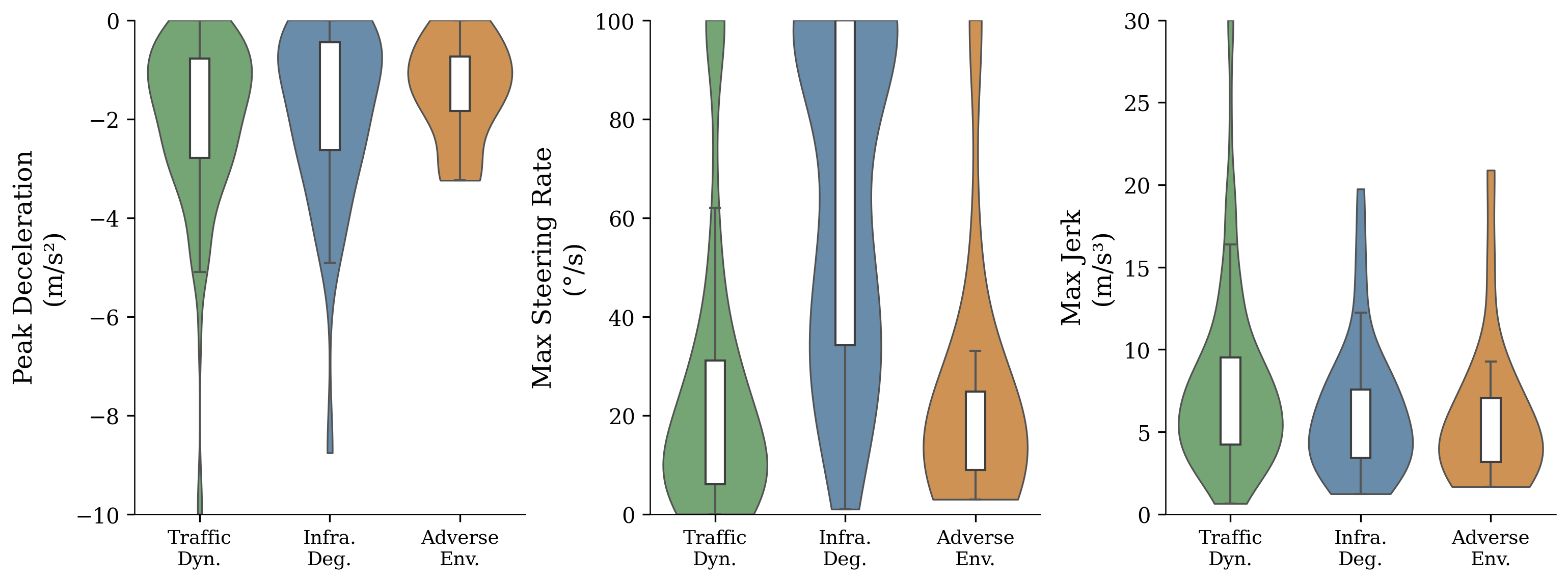}
    \caption{Post-takeover kinematic metrics across semantic archetypes. }
    \label{fig:kinematic_signatures}
\end{figure}

\textbf{Infrastructure degradation correlates with high lateral intervention.} Takeovers triggered by missing lane lines or construction exhibit a median steering rate of 71.8$^\circ$/s, substantially higher than other archetypes. Longitudinal deceleration remains moderate (-1.37\,$\text{m/s}^2$). This suggests that when lateral guidance fails due to degraded markings, drivers primarily execute sharp steering corrections to maintain lane position without requiring heavy braking.

\textbf{Traffic dynamics induces longitudinal instability.} Encounters with slow or suddenly braking vehicles result in the highest median jerk (6.69\,$\text{m/s}^3$) and peak deceleration (-1.61\,$\text{m/s}^2$), while steering rates remain comparatively low (15.9$^\circ$/s). This indicates that lagging ADAS longitudinal responses in dynamic traffic force drivers to execute abrupt braking maneuvers to avoid collisions.

\textbf{Adverse environments are associated with preemptive interventions.} Takeovers in visually degraded conditions demonstrate the lowest kinematic severity across all metrics (median jerk 4.28\,$\text{m/s}^3$, deceleration -1.22\,$\text{m/s}^2$, steer rate 13.1$^\circ$/s). This implies a risk compensation effect: in challenging environments, drivers tend to reclaim control earlier and apply smoother inputs.

These results demonstrate that distinct environmental hazards elicit specific kinematic responses. The correlation between visual semantics and physical intervention severity highlights the value of multimodal perception, providing a basis for early warning systems and parameterized safety models discussed in the following section.

\section{Discussion}
\label{sec:discussion}

\subsection{Temporal Advantage Over Traditional Collision Avoidance}
\label{sec:early_warning}

\begin{figure}[htbp]
    \centering
    \includegraphics[width=\columnwidth]{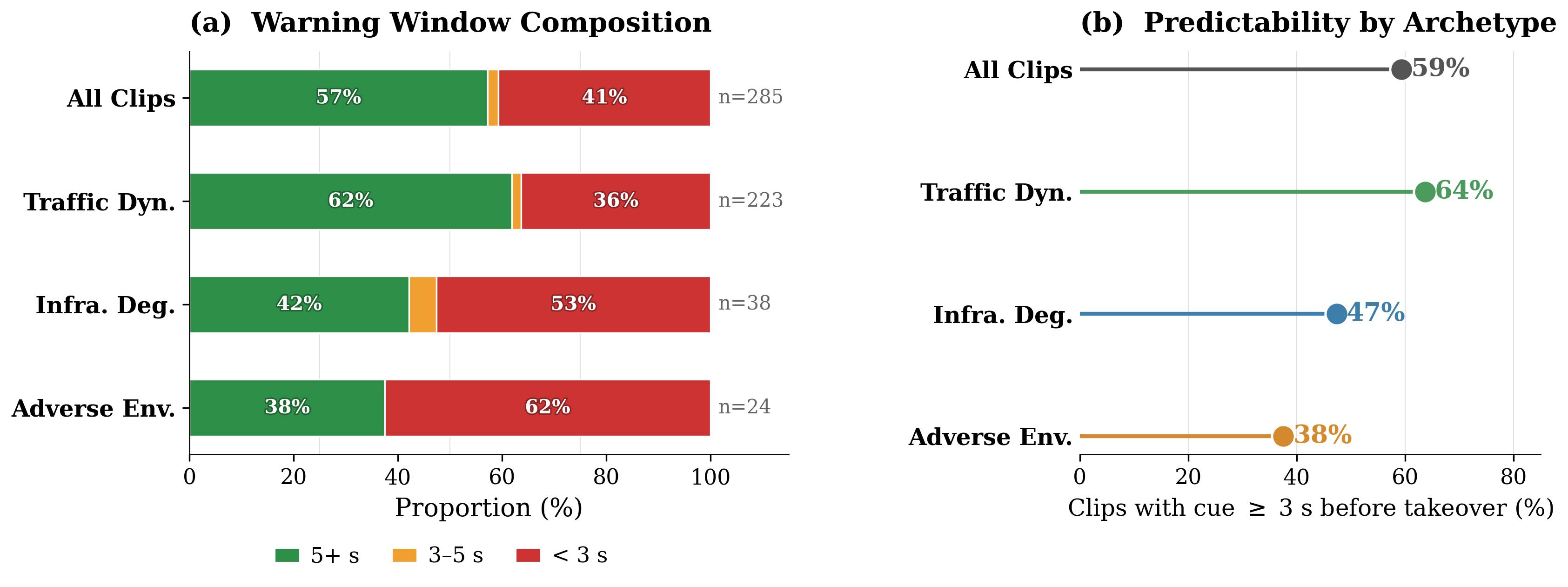}
    \caption{Temporal advantage of VLM-based hazard detection. Left: Distribution of the earliest actionable visual cues. Right: Predictability ($\ge 3$\,s warning) stratified by hazard archetypes.}
    \label{fig:early_warning_advantage}
\end{figure}

Traditional collision avoidance systems (FCW, AEB) rely on late-stage kinematic thresholds (TTC, THW), frequently forcing abrupt driver interventions. To assess the viability of proactive alerting, we quantified the temporal advantage of VLM-extracted semantic cues across the 285 critical events. As shown in Fig.~\ref{fig:early_warning_advantage}, 59.3\% of critical clips contain actionable visual cues at least 3 seconds before the takeover, with 57.2\% discernible as early as $T-5$\,s. 

\begin{figure}[htbp]
    \centering
    \includegraphics[width=\columnwidth]{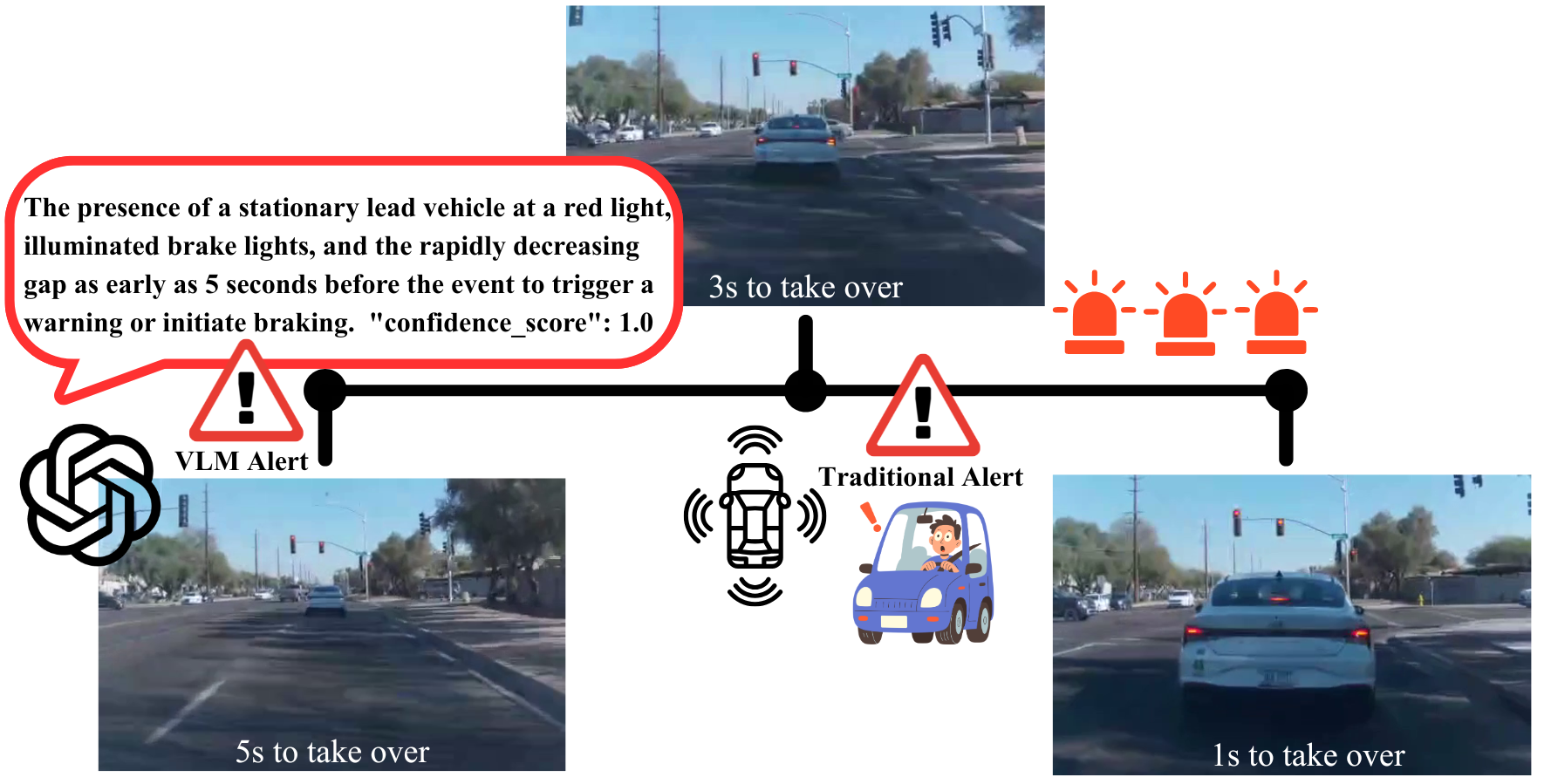}
    \caption{Comparison of VLM-based early warning and traditional systems. The VLM detects the red light and brake lights at $T-5$\,s, establishing hazard awareness well before kinematic thresholds trigger near $T-2$\,s.}
    \label{fig:early_warning_demo}
\end{figure}

This temporal predictability heavily depends on the hazard archetype. Traffic Dynamics is highly anticipatable (63.7\% predictability) due to salient cues like leading vehicle brake lights. Infrastructure Degradation offers moderate predictability (47.4\%), as faded lines often manifest only at close range. Adverse Environments represent the primary limitation of camera-based perception, yielding the lowest predictability (37.5\%).

Fig.~\ref{fig:early_warning_demo} illustrates this advantage in a traffic light scenario. While conventional kinematic systems trigger near $T-2$\,s as TTC collapses, the VLM identifies the red traffic light and the lead vehicle's brake lights at $T-5$\,s. Integrating such predictive semantics enables graduated, early warnings, eliminating the need for aggressive evasive maneuvers.

\subsection{Limitations and Future Work}
\label{sec:limitations}
Our current takeover intent partition and hazard labeling are intentionally lightweight, which may miss fine-grained intent and context variations. Future work will refine these labels with targeted human annotation and larger-scale VLM-assisted tagging, and use the improved labels to design and validate semantics-aware early-warning policies that fuse scene cues with kinematic risk metrics.

\bibliographystyle{IEEEtran}
\bibliography{references}
\end{document}